\documentclass[mathleft
]{an}
\usepackage{graphicx}
\usepackage{rotating}
\usepackage{lscape}
\usepackage{longtable}
\usepackage{times}
\begin{document}

\Pagespan{789}{}
\Yearpublication{2011}%
\Yearsubmission{2010}%
\Month{11}%
\Volume{999}%
\Issue{88}%

\title{A solar super-flare as cause for the $^{14}$C variation in AD 774/5~?}

\author{R. Neuh\"auser\inst{1} \thanks{Corresponding author: \email{rne@astro.uni-jena.de}}
\and V.V. Hambaryan\inst{1}
}

\titlerunning{Solar super-flare in AD 774/5~?}
\authorrunning{Neuh\"auser \& Hambaryan}

\institute{
Astrophysikalisches Institut und Universit\"ats-Sternwarte, FSU Jena,
Schillerg\"a\ss chen 2-3, D-07745 Jena, Germany
}

\received{2014}
\accepted{May 2014}
\publonline{ }


\abstract{We present further considerations regarding the strong $^{14}$C variation in AD 774/5.
For its cause, either a solar super-flare or a short Gamma-Ray Burst were suggested.
We show that all kinds of stellar or neutron star flares would be too weak for the observed
energy input at Earth in AD 774/5.
Even though Maehara et al. (2012) present two super-flares with $\sim 10^{35}$ erg of presumably solar-type
stars, we would like to caution: These two stars are poorly studied and may well be close binaries, 
and/or having a M-type dwarf companion, and/or may be much younger and/or much more magnetic than the Sun -
in any such case, they might not be true solar analog stars.
From the frequency of large stellar flares averaged over all stellar activity phases 
(maybe obtained only during grand activity maxima), 
one can derive (a limit of) the probability for a large solar flare at a random time of normal activity:
We find the probability for one flare within 3000 years to be possibly as low
as $0.3$ to $0.008$ considering the full $1~\sigma$ error range.
Given the energy estimate in Miyake et al. (2012) for the AD 774/5 event, 
it would need to be $\sim 2000$ stronger than the Carrington event as solar super-flare.
If the AD 774/5 event as solar flare would be beamed (to an angle of only $\sim 24^{\circ}$), 
100 times lower 
energy would be needed. 
A new AD 774/5 energy estimate by Usoskin et al. (2013) with a different carbon cycle model,
yielding 4 or 6 time lower $^{14}$C production, predicts 4-6 times less energy.
If both reductions are applied, the AD 774/5 event would need to be only $\sim 4$ times stronger 
than the Carrington event in 1859 (if both had similar spectra).
However, neither $^{14}$C nor $^{10}$Be peaks were found around AD 1859. 
Hence, the AD 774/5 event (as solar flare) either was not beamed that strongly,
and/or it would have been much more than 4-6 times stronger than Carrington,
and/or the lower energy estimate (Usoskin et al. 2013) is not correct,
and/or such solar flares cannot form (enough) $^{14}$C and $^{10}$Be. 
The 1956 solar energetic particle event was followed by a small {\em de}crease in directly observed cosmic rays.
We conclude that large solar super-flares remain very unlikely as the cause for the $^{14}$C increase in AD 774/5.
}

\maketitle

\section{Introduction: $^{14}$C variation in AD 774/5}

A significant variation (by $7.2~\sigma$) was detected in the isotope ratio of $^{14}$C to $^{12}$C  
in two Japanese cedar trees (Cryptomeria japonica), an increase by $+1.2~\%$ in the year AD 774/5,
which was followed by a subsequent decrease over $\sim 10$ to 20 yr; the increase in $^{14}$C is consistent with an 
increase of this isotope in IntCal98 from European and northern American trees, which are available with 5 to 10 yr 
time resolution (Miyake et al. 2012, henceforth M12, Stuiver et al. 1998a). 
If the $^{14}$C was deposited within $\le 1$ yr, best consistent with the atmospheric deposition model,
the increase corresponds to $\ge 19 \pm 4$ atoms cm$^{-2}$ s$^{-1}$ (M12).
$^{14}$C atoms are produced in the atmosphere either be high energetic particles (e.g. protons) or $\gamma$-rays in
a nucleonic/electromagnetic muon cascade.
This $^{14}$C increase requires an energy of $7 \times 10^{24}$ erg at Earth, 
if the radioisotopes were formed due to incoming $\gamma$-photons above 10 MeV
with a supernova (SN)-like spectrum with a power-law index of $-2.5$ (M12).
With a different carbon cycle model, Usoskin et al. (2013, henceforth U13) need
4 to 6 times less $^{14}$C production and, hence, 4 to 6 times less energy.
Furthermore, a $30~\%$ increase in $^{10}$Be flux was observed for the decade around AD 785 
on Antarctica with 10-yr time resolution (Horiuchi et al. 2008), 
maybe produced in the same event as the $^{14}$C; $^{10}$Be data have lower timing precision.
In case that protons from a solar flare or a solar energetic particle event
(such as a Solar Proton Event (SPE) or a Coronal Mass Ejection) 
would have been the original source for the $^{14}$C and $^{10}$Be
production, one would need a proton energy of $8 \times 10^{25}$ erg at Earth or
$2 \times 10^{35}$ erg at the Sun (M12); or 4 to 6 times less energy (U13).

Such high energy estimates would be neeeded for ions at the Sun. However, solar energetic particles
are not accelerated at the Sun, but in interplanetery space. Thus, one may envisage acceleration close
to Earth, where much less protons could achieve the same result. A possible scenario is the collision
of two interplanetary shocks. This would imply that the solar super-flare interpretation
would not fit within the current knowledge of solar and stellar flares.

\subsection{Possible causes}

While M12 already argued that solar and stellar flares as well as a normal unreddened SN
explosions are unlikely as cause for this event, Hambaryan \& Neuh\"auser (2013, henceforth HN13) also 
found reddened SNe (with the typical $\sim 10^{51}$ erg total energy, $1~\%$ of it as $\gamma$-rays) 
to be very unlikely to explain the AD 774/5 event; the SN would need to be at $\sim 124$ pc only (HN13). 
If the (small) $^{14}$C increase observed about 3 yr after
AD 1006 (Damon et al. 1995) was due to SN 1006, the brightest optical SN as seen from Earth in the last millennia,
then $10^{50}$ erg $\gamma$-ray energy would have been required (Damon et al. 1995), so that the AD 774/5
event could still be due to a similar (over-luminous) SN, but brighter (in $\gamma$) and/or closer than SN 1006;
then, it should have been observable, but there are no such historic records;
a young SN remnant at, e.g., $\sim 1$ kpc distance would also be detectable, but might still be undetected.
Menjo et al. (2005) and Miyake et al. (2013, henceforth M13) also see an increase in $^{14}$C around AD 1009 by a few p.m.,
but the significance was not as large as in Damon et al. (1995). Menjo et al. (2005) considered 
whether the increase was solely or mostly due to solar activity. 

The nova or SN candidate listed as {\em Hye Sung} (an ancient Korean comet name) in Chu (1968) for 
AD 776 Jun 1-30 in Tau-Aur (from the Korean Lee Dynasty chronology during the reign of He Gong Sinla) 
is more likely a comet or nova (observed for only 30 days) - and maybe also too late for the AD 774/5 event,
if both are dated correctly.

HN13 also showed that all observables of the AD 774/5 event are consistent with a short gamma-ray burst (GRB),
while a long GRB would not yield the correct $^{14}$C and $^{10}$Be production ratio (HN13).
More recently, Pavlov et al. (2013 a,b) confirmed the approximate estimates by HN13 with more precise
calculations using GEANT and argued that a Galactic {\em long} GRB is also not inconsistent with
the AD 774/5 observables.

Several further considerations were also published recently: \\
Allen (2012) suggested 
that a report in the Anglo-Saxon Chronicle
({\em This year also appeared in the heavens a red crucifix, after sunset}),
presumably for AD 774
may be a reddened SN; however, it is clear
that a SN cannot be observed as resolved or extended object ({\em cross}) and that
a nearby SN remnant should be observable anyway behind extinction in $\gamma$- or X-rays;
the frequent sightings of a {\em cross seen at sky} in medieval times
can easily be explained as well-known phenomenon of parhelion or paraselene (Neuh\"auser \& Neuh\"auser 2014a). 
It was also shown that the above {\em red cross} was observed in AD 776, not AD 774/5 (e.g. Gibbons \& Werner 2012). 

Eichler \& Mordecai (2012) argued that a large solar flare or proton event cannot explain
the event (as also argued in M12), but an impact of a massive comet onto the Sun may be able to explain the energetics. 

A few authors still consider a solar super-flare:
The solar activity was reconstructed for past centuries and millennia from sunspot and aurora observations,
which may, however, be incomplete, biased, and inhomogeneous; hence, radionuclide archives on Earth
were used: The larger the solar activity, the larger the number of sunspots and aurorae observed,
and the stronger the solar wind and, hence, the smaller the incoming cosmic-rays, hence, a {\em decrease} in the
production ratio of radionuclids (see e.g. the recent review by Usoskin 2013). 
However, $^{14}$C from tree rings and $^{10}$Be from ice cannot give a time resolution of better than 1 yr,
while sunspot and aurora observations can indicate solar activity changes on time-scales of days.

Usoskin \& Kovaltsov (2012), Melott \& Thomas (2012), Thomas et al. (2013), and U13 suggested 
that a (possibly beamed) large SPE could be the cause for the strong sudden {\em increase} in radionuclids.
Neither Eichler \& Mordecai (2012), Melott \& Thomas (2012) nor Thomas et al. (2013) 
considered whether the suggested events can explain the differential $^{14}$C to $^{10}$Be production ratio 
observed to be $\ge 270 \pm 140$ (HN13).\footnote{HN13 used the M12 data to obtain this rough,
approximative and conservative estimate, intended for disentangling between cosmic-ray/SPE and $\gamma$-ray 
scenario in the simplest way - corresponding to the limiting case, i.e. for a given energy range (20-60 MeV),
energy distribution of secondary neutrons can be considered almost constant.}
According to Usoskin et al. (2006) and Usoskin \& Kovaltsov (2012) a SPE can result in a $^{14}$C to $^{10}$Be production ratio
of only $\sim 25$ to 39, i.e. too small (compared to the observed ratio of $\ge 270 \pm 140$, HN13). 
Masarik \& Ready (1995) and Usoskin et al. (2006) concluded that the effect of $^{14}$C production due to solar particles is
negligible (less than $1\%$ on average) for both with cascades and without cascades, confirmed again by Usoskin \& Kovaltsov (2010).
M12 excluded a solar flare partly based on the fact that such flares are
not hard enough to explain the differential $^{14}$C to $^{10}$Be production ratio. 

Most recently, Cliver et al. (2014) argued that - by comparing energetics and spectrum of the hardest solar
SPE in the last century (SPE 1956) with the AD 774/5 event -
the AD 774/5 event was most probable not a solar super-flare (and the Sun was in a low-activity state at
around AD 774, as claimed by Cliver et al. 2014).

Even more recently, Liu et al. (2014) suggested that a presumable comet impact on Earth on AD 773 Jan 17 (presumably
observed in China) was responsible for the input of large amounts of $^{14}$C (and possibly $^{10}$Be) to the
Earth atmosphere at that time. They resolved $^{14}$C with a time-resolution of only 2 weeks in corals
in the South China Sea for two years lying somewhere around AD $783 \pm 14$ ($^{230}$Th dating),
i.e. possibly near AD 773/4. Their coral $^{14}$C fluctuates by $\pm 25$ pm within 20 years (with lower
time-resolution) and by $\pm 30$ pm within the two years with higher time-resolution.
Liu et al. (2014) claim that the first rise in $^{14}$C seen in their data correlates with the 
sighting of a {\em comet} collision {\em with the Earth atmosphere from the constellation of Orion on
17 Jan 773 with coma stretched across the whole sky and disappeared within one day, 
with dust rain in the daytime}, 
a presumable sighting from the Chinese Tang dynasty. \\
For the dating given in Liu et al. (2014), i.e. $^{14}$C input to the atmosphere on 13 Jan 773,
it would be surprising that the $^{14}$C increase was first seen in coral, but one year later in trees. \\
There was no rise in $^{14}$C nor $^{10}$Be after the Tunguska event on AD 1908 Jun 30 
(see our Fig. 1)\footnote{The (anti-)correlation between sunspot phases and $^{14}$C works well after having 
shifted the $^{14}$C values backwards by $\sim 2$ yr given the carbon cycle (yearly incorporation into trees peaks 
after 2-3 yr); when we shift backwards the $^{14}$C values (in Figs. 1, 2, and 3) by 2 yr from the published (integer value) years,
we effectively shift by $\sim 2.5$ yr, because tree rings are mainly formed during summer
(all trees used are northern hemisphere trees).}, 
when a comet or asteroid hit the Earth atmosphere with strong devastations in Russia
(also found by Melott et al. 2010).
Given that there are apparently no devastations known or related to an event on AD 773 Jan 17,
one might conclude that the AD 773 Jan 17 event was smaller (lower mass object) than the Tunguska event -
in such a case, we would not expect a rise in $^{14}$C if such an object would have hit Earth in AD 773. \\
Overholt \& Melott (2013), Usoskin \& Kovaltsov (2014), and Melott (2014) show that such a large amount of $^{14}$C cannot be deposited
to Earth with a comet nor asteriod, unless for a very large body, which would cause very strong devastations.
Usoskin \& Kovaltsov (2014) do not consider whether the Liu et al. (2014) $^{14}$C data with very high time-resolution
are consistent with a solar flare: The Liu et al. (2014) data do show a strong increase by some 10 p.m. within a short time intervall
(in their half-annual data) and also three strong jumps by $\sim 30$ to 55 p.m. within 2-4 weeks in their bi-weekly data, 
all of them sometime during AD $783 \pm 14$ given by their $^{230}$Th dating,
but there is another jump by $\sim 15$ p.m. within one year some 12 years later (again in the half-annual data).
Could this be due to normal solar activity variation~? \\
Chapman et al. (2014) show that the original Chinese texts about this comet just report about
a very normal comet observed in China on or since AD 773 Jan 17 (also observed in Japan on or since Jan 20),
and that the material presented in Liu et al. (2014) is misleading: There is no evidence for a collision 
of a comet with Earth. 

U13 argued that M12 overestimated the number of produced $^{14}$C atoms and,
hence, the energy input at Earth by a factor of 4 to 6 due to their model of incorporation of
$^{14}$C atoms into tree rings. Given that the $^{10}$Be production should remain the same, 
the differential $^{14}$C to $^{10}$Be production ratio would be only $\ge 54 \pm 30$ 
(scaled from HN13) and, hence, 
more compatible with expectations from Usoskin et al. (2006) and Usoskin \& Kovaltsov (2012).

\begin{figure}
\begin{center}
{\includegraphics[angle=270,width=8cm]{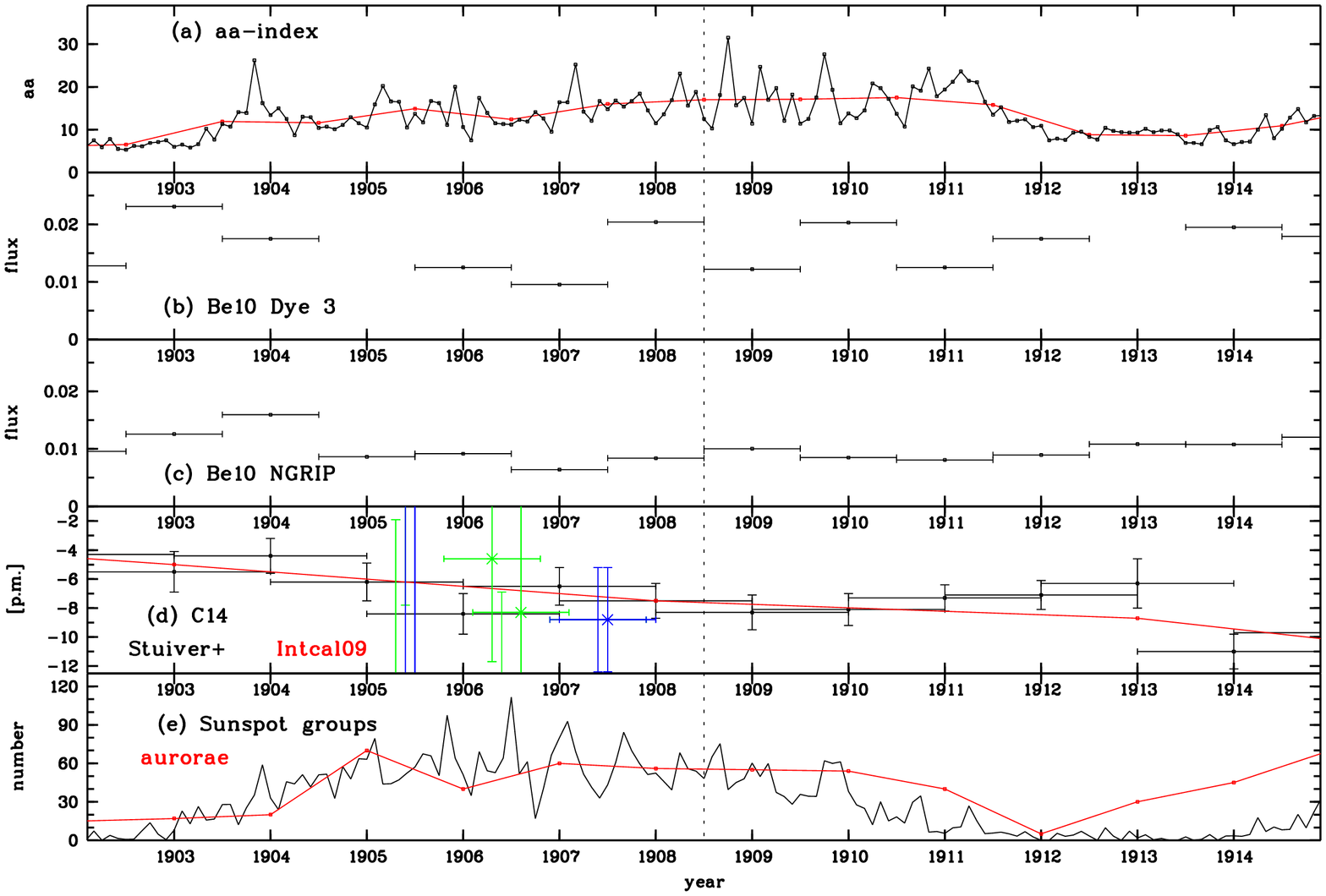}}
\caption{Solar activity proxies around the AD 1908 Jun 30 Tunguska event (1902-1915):
(a) aa-index, both monthly in black and yearly in red.
(b) $^{10}$Be flux in atoms cm$^{-2}$ s$^{-1}$ with 1-yr time resolution (Dye-3 from Beer et al. 1990).
(c) $^{10}$Be flux in atoms cm$^{-2}$ s$^{-1}$ with 1-yr time resolution (NGRIP from Berggren et al. 2009ab).
(d) $^{14}$C data in p.m. from IntCal09 with 5-yr time resolution in red (Reimer et al. 2009),
with 1-yr time resolution in black (Stuiver et al. 1998b), 
and from Yonenobu \& Takenaka (1998) in blue 
(from trees in Tunguska) and green (trees from Japan) - all plotted 2.5 yr ahead of their
measurement time (in the northern summer of a tree ring year $x$, hence incorporated at epoch $x.5$), 
because they were produced in the atmosphere $\sim 2.5$ yr earlier due to the carbon cycle.
The small rise seen in $^{14}$C from 1909 to 1913 is just the normal modulation due to the Schwabe cycle.
as sunspots decrease to their minimum in 1913.6.
(e) Monthly sunspot group numbers in black (Hoyt \& Schatten 1998) and yearly aurorae from Legrand \& Simon (1987) in red.
The dotted line indicates the date of the Tunguska event: AD 1908 Jun 30.
There is no signal in $^{10}$Be nor $^{14}$C. 
We agree with Melott et al. (2010) by concluding that the Tunguska event did not bring in any detectable $^{10}$Be nor $^{14}$C.
(Yonenobu \& Takenaka (1998) found a small overabundance found in trees grown in 1909, i.e. they did not
take the carbon cycle into consideration (that it takes one to few years until the event is seen in trees),
and their presumable effect was found to be a purely botanical effect by Suess (1965); 
we see their Tunguska tree data as blue dots (and their Japanese controll sample as green dots)
in panel (d), all plotted as usual 2.5 yr ahead of their measurements; 
we conclude that their data are consistent with the other data plotted within their error bars.)}
\end{center}
\end{figure}

While M12 - not making an assumption on the cause of the event - used the long-term average 
of 2.05 $^{14}$C atoms cm$^{-2}$ s$^{-1}$ as cosmic-ray background (average over all solar activity phases), 
U13 used 1.6 $^{14}$C atoms cm$^{-2}$ s$^{-1}$ as cosmic-ray background, 
also as long-term average over all solar activity phases.
The most recent study gave a cosmic-ray background production rate of $1.6^{+0.6}_{-0.3}$ atoms cm$^{-2}$ s$^{-1}$
for a constant geomagnetic field (Roth \& Joos 2013).

\subsection{The event rate}

Eichler \& Mordecai (2012), Melott \& Thomas (2012), and Thomas et al. (2013) all consider the rate
of the $^{14}$C event observed by M12 to be one in $\sim 1250$ yr, just because the event in AD 774/5
was about 1250 yr ago. For the last 3000 yr, exactly one significant ($\ge 3 \sigma$), rapid (within 1 yr),
and large (larger than 3 p.m.) increase was observed in IntCal98 data, namely in the AD 770ies (M12).
According to M12, there are three such significant ($\ge 3 \sigma$) increases
observed with 10-yr time resolution, which were then all observed with time resolution 
of 1 to 2 yr (one of them BC, the two others in the AD 770ies and 1790ies).
Two of the three events were found to be events on time-scales of several years, while only one of the three
events was not only significant ($\ge 3 \sigma$) and large (larger than 3 p.m.), but also rapid,
i.e. an increase within 1 yr, namely in AD 774/5 (M12). Hence, even though the high time-resolution data are not
available for 3000 yr, it is clear that there was only one such strong (as strong as AD 774/5) event 
within the last 3000 yr\footnote{
If there would be another large $^{14}$C spike as in AD 774/5 (to be found
with high time-resolution within the last 3000 yr), such data would need to be consistent with IntCal data, which is 
possible only, if there would also be a rapid decrease within the same (Intcal) time step; this is (almost) impossible.
As mentioned above, the variation found with high time-resolution in AD 774/5 was seen before in IntCal (M12).}
(so that a consideration of a rate of once in 1250 yr is not justified).
Eichler \& Mordecai (2012), Melott \& Thomas (2012), and Thomas et al. (2013) then argue
that the rate of massive comet impacts on the Sun and also of large solar flares, respectively, with sufficient 
energy to explain the AD 774/5 event, would just be consistent with once in 1250 yr.

The IntCal $^{14}$C data with 5- to 10-yr resolution are available for the last 11,000 yr.
According to Usoskin \& Kovaltsov (2012), the IntCal $^{14}$C data (Stuiver et al. 1998a) show 
even only one such significant ($\ge 3 \sigma$) large (larger than 3 p.m.) increase
within the last 11,000 yr. Hence, we have to consider a rate of one strong event in 11,000 yr.
Hence, if the rate of massive comet impacts on the Sun and also of large solar flares with
sufficient energy would be one in 1250 yr, then we should have observed several more
such strong events in the last 11,000 yr, which is not the case. A rate of one such super-flare 
every 1250 yr can be excluded from the 11,000 yr $^{14}$C data base with high confidence.

Recently, M13 presented more $^{14}$C data with 1 yr
time resolution for AD 822 to 1021 with one more rapid increa- se\footnote{Originally dated 1 year earlier
in AD 992/3, but corrected by a Nature Corrigendum in 2013 November due to a mis-count of the tree rings.} 
from AD 993 to 994 with 
a slow decrease,
namely an increase by 9.1 p.m. with $5.1~\sigma$ significance,
i.e. smaller than in AD 774/5.
The IntCal09 data show this event as an increase by 3 p.m. from AD 980 to 995 (M13).
This $^{14}$C increase at around AD 993/4 was already detected in data with 2 yr time resolution
published by Menjo et al. (2005) at slightly lower amplitude.
M13 show that there is also a $^{10}$Be increase from AD 985 to 995 in the 
Dome Fuji data (Horiuchi et al. 2008)\footnote{M13 argue that the AD 774/5 and AD 993/4 events show a similar
$^{14}$C to $^{10}$Be production ratio.
However, the observation that in both events the $^{14}$C and
$^{10}$Be peaks appear to be at least in the same decade is obtained only after the $^{10}$Be age data were
corrected by matching the $^{14}$C pattern (Horiuchi et al. 2008); hence, it is not yet proven that both radioisotopes
were produced at the same time. In the GRIP $^{10}$Be data, the increase in $^{10}$Be is 15 yr earlier than in Dome Fuji data,
and there is no peak at AD 993/4, the curve is flat from AD 900 to 1000. 
We also note that $^{10}$Be data have much lower time resolution and less time precision.}.
The rate of strong events like the AD 774/5 event with an increase of 12 to 15 p.m. within 1 yr remains as before,
as such strong events would be visible in data with 10-yr time resolution;
the rate of events at least as strong as the AD 993/4 event, which are detectable only
with data with 1- or 2-yr time resolution (not in data with 10-yr time resolution), is 
two events in 1130 yr, as $^{14}$C data with 1 to 2 yr time resolution are available for 1130 yr:
From AD 600-1021 (M12, M13, Miyake et al. 2014), then from AD 992-1150 (Damon et al. 1995, 1998, Damon \& Peristykh 2000, Menjo et al. 2005), 
then from AD 1374-1745 (Miyahara et al. 2004, 2006, 2007, 2010), and then from AD 1510-1954 (Stuiver et al. 1998b);
for the time since about AD 1900, $^{14}$C data are affected by the Suess effect, and since about AD 1954 also by the bomb effect,
so that we cannot consider the data after AD 1954 here.
(In M13, also a few additional data points from tree A for the time between AD 770 and 800 were taken.)

In this article, we extend the discussion about solar and stellar flares.
First, we briefly show that flares from neutron stars (Sect. 2.1) or stars other
than the Sun (Sect. 2.2) cannot explain the AD 774/5 event due to limited energetics.
Then, we estimate the general probability of a large solar flare with the neccessary energetics (Sect. 3).
We then discuss the probability for a very large solar super-flare in AD 774/5 
and summarize our results in Sect. 4.

\section{A stellar flare~?}

Solar and stellar flares (or SPEs) were found unlikely based on the argument that such strong flares
were never observed on the Sun and that the strongest flares observed on other stars 
were neither strong nor hard enough to produce the differential $^{14}$C to $^{10}$Be production ratios
as observed in AD 774/5 (M12). The first argument was challenged by Melott \& Thomas (2012),
the latter by Usoskin \& Kovaltsov (2012) and U13. We extend this discussion here.

As in HN13, we can estimate the distance of an event to be able to produce the observed energy input 
at Earth ($E_{\rm obs}$) in the following way:
The ratio between the energy emitted by an event ($E_{\rm event}$)
spread homogeneously into the total surface area of a spherical shell around
the event ($4 \cdot \pi \cdot d^{2}$ with distance $d$ from the event to Earth) and the
energy $E_{\rm obs}$ observed at Earth is equal to the ratio between the surface area of that sphere and
the Earth solid angle $\pi \cdot R^{2}$ (with Earth radius $R = 6378$ km):
\begin{equation}
\frac{E_{\rm event}}{E_{\rm obs}} = \frac{4 \cdot \pi \cdot d^{2}}{\pi \cdot R^{2}}
\end{equation}

According to M12, the energy observed at Earth is either $(7.0 \pm 1.5) \times 10^{24}$ erg if the
radioisotopes were produced by $\gamma$-rays above 10 MeV with a SN-like spectrum with a power-law index of $-2.5$,
or $(8.0 \pm 1.7) \times 10^{25}$ erg at Earth (i.e. $(2.0 \pm 0.4) \times 10^{35}$ erg at the Sun)
if produced by protons.
According to U13, the radiocarbon production (and, hence, also the energy input to the Earth atmosphere) 
was 4 to 6 times lower than given in M12. Then, the Earth was hit by either $(1.4 \pm 0.4) \times 10^{24}$ erg 
if the radioisotopes were produced by $\gamma$-rays (again above 10 MeV with a SN-like spectrum with a power-law index 
of $-2.5$ just scaled down by a factor of 4 to 6),
or by $(0.4 \pm 0.1) \times 10^{35}$ erg at the Sun if the observed $^{14}$C excess was produced by protons.

Melott \& Thomas (2012) argue that solar flares
(in particular SPEs) can be beamed with observed angles of $24^{\circ}$ to $72^{\circ}$ 
(Bothmer \& Zhukov 2007). Given that an angle of $24^{\circ}$ corresponds to 0.01 of the
total surface area of a sphere, the 
energy of a stellar or solar flare, 
in order to be consistent with the AD 774/5 event, can be reduced by a factor of up to 100.
For radioisotope production by protons, the solar flare would then need to have an 
energy of $(2.0 \pm 0.4) \times 10^{33}$ erg at the Sun
for the M12 $^{14}$C and energy estimate,
or only $(0.4 \pm 0.1) \times 10^{33}$ erg at the Sun for the U13 $^{14}$C and energy estimate.

As mentioned above, such high energy estimates would be neccessary for ions at the Sun, but solar energetic particles (SEP)
are closer to Earth in interplanetery space. For the moment, no theoretical and/or numerical model treats SEP acceleration 
and transport near its full complexity. An interesting point in the model by Zank et al. (2000) is that, for extremely strong shocks, 
particle energies of the order of 1 GeV can be achieved when the shock is still close to the Sun. 
As the shock propagates outward, the maximum accelerated particle energy decreases sharply. 
Other shock acceleration models (Berezhko et al. 2001) also suggest the possibility that 1 GeV protons 
can be accelerated when extremely strong shocks are close to the Sun (within 3 solar radii). 
Comparisons of these models of particle shock-acceleration with specific observations have not yet been reported.
Given that even the hardest SEP of the last century (SEP 1956) was found not to be hard and strong enough in comparison
with the AD 774/5 event (Cliver et al. 2014), the acceleration of high-energy particles in AD 774/5 
(if it would have been a flare) would be expected to be closer to Sun than to Earth.

Protons or other charged particles from any distant event within several pc from outside the
solar system would be dispersed by the galactic magnetic field:
If produced and emitted at a distance $d$, they will arrive at Earth 
after the diffusion time of $\sim 3 \cdot$ (d[pc])$^{2}$ yr due
to the inhomogeneous Galactic magnetic field (Laster 1968).
Hence, even from a small distance of only $\ge 2$ pc, they will be dispersed over $\ge 4$ years,
which seems inconsistent with the $^{14}$C data (M12).
Therefore, for the AD 774/5 $^{14}$C increase, we do not need to consider radioisotope production by protons
from stars other than the Sun or neutron stars
(but the production by $\gamma$-rays would still need to be considered).

\subsection{Strong flares on Neutron Stars}

We extend here the brief discussion of neutron star flares given in HN13.

Large flares on neutron stars such as magnetars, i.e. soft $\gamma$-ray repeaters (SGR) 
or Anomalous X-ray Pulsars (AXPs) were not considered in M12. 
The largest magnetar flare observed so far was the X- and $\gamma$-ray flare of SGR 1806-20 on 2004 Dec 27,
it had a peak energy output of $3.7 \pm 0.9 \cdot 10^{46}$ erg/s 
at the previously assumed distance of 15 kpc (Hurley et al. 2005); 
this corresponds to $E_{\rm event} = 2 \cdot 10^{46}$ erg at the revised distance, $8.7 \pm 1.7$ kpc (Bibby et al. 2008).  
If the AD 774/5 event would have been due to a spherical neutron star flare (expanding fireball) with such an energy, 
we can estimate the required distance $d$ from Equ. 1 to be $d \simeq 5.5$ pc. 
There is no neutron star known at or within this distance -
neither in the ATNF catalog\footnote{www.atnf.csiro.au/people/pulsar/psrcat/} 
(Manchester et al. 2005) nor in the McGill magnetar 
catalog\footnote{www.physics.mcgill.ca/~pulsar/magnetar/main.html} of SGRs and AXPs.

The above arguement was already given in HN13, let us now extend the discussion.

SGR 1806-20 has a magnetic field strength of $B = 2.4 \times 10^{15}$ G, and the flare energy $E$
is thought to be proportional to the square of the magnetic field strength ($E \sim B^{2}$).
Hence, a magnetar with $10^{16}$ G dipole field could produce an event with $10^{48}$ erg (Hurley et al. 2005).
Then, the distance $d$ of such a neutron star would have to be $d \simeq 39$ pc (from Equ. 1) -
again, there is no such neutron star known within that small distance.
A magnetar at such a small distance would be known, as it would have been detected 
by the ROSAT all-sky X-ray survey: With a typical persistent bolometric luminosity 
of $\sim 0.025$ to $1.6 \cdot 10^{35}$ erg/s (mostly in X-rays) with typical spectral components 
of magnetars (blackbody with peak energy $k \cdot T = 0.4$ keV and power-law index $\sim 3$,
as observed and given in the McGill SGR/AXP catalog, footnote 7, 
we would expect 150 to 800000 counts per second in the ROSAT energy band (0.1 to 2.4 keV) at 
distances from 10 to 100 pc, i.e. very easily detectable. 
Hence, neutron star flares were also found to be very unlikely to
be the cause of the AD 774/5 event (HN13).

Even if neutron star flares would be beamed, this would not change the conclusion:
If neutron stars flares are usually beamed (by some typical beaming angles),
than the X- and $\gamma$-ray flare of SGR 1806-20 on 2004 Dec 27 would have been beamed that way, too.
Our scaling from an unbeamed flare (of SGR 1806-20 on 2004 Dec 27)
to a possible flare in AD 774/5 would be the same, regardless of whether we
assume that both flares would have been beamed (in a similar way) or that both flares would have been unbeamed.
In addition, giant neutron star flares are seen as spherically expanding relativistic plasma
radiating as a thermal fireball trapped in the magnetosphere (Mereghetti 2008), hence spherical.

It is also extremely unlikely that a possible neutron star as source for the AD 774/5 event (due to a strong SGR-like flare)
would have been an active AXP or SGR in AD 774/5, but became inactive (and also undetected) since then. The space density 
of known AXPs and SGRs is very small: 24 magnetars are known in our half of the Galaxy, i.e. none observed behind the Galactic Center
and excluding those few in other galaxies (McGill magnetar catalog, see footnote 7), which gives a space density of
$\sim 0.1$~kpc$^{-3}$, so that the probability to expect one within $d \simeq 39$ pc (see above) is extremly low ($\sim 10^{-5}$).

We would like to note in passing that, even though of our distance estimate of $\sim 124$ pc (HN13)
for the AD 774/5 event as normal SN, and even though there is a neutron star known at that
distance, namely RXJ1856.5-3754 (Walter et al. 1996) at $123 \pm 13$ pc (Walter et al. 2010),
this object is not the counterpart of the AD 774/5 event,
because this neutron star is much older: Since its blackbody emission peaks in the soft X-ray regime at
$\sim 63$ eV (Burwitz et al. 2001, 2003, Hohle et al. 2012),
it must have cooled down for several 100 kyr according to normal neutron star cooling curves;
its motion points back to the Upper Scorpius OB association
(Walter 2001, Walter \& Lattimer 2002), where it might have formed in a SN some 0.4 to 0.5 Myr ago
(Tetzlaff et al. 2010, 2011). The situation is similar for the other isolated thermal neutron stars called
Magnificent Seven, which are all soft X-ray emitters (e.g. Neuh\"auser et al. 2011), probably all similar to
RXJ1856.5-3754, i.e. a few 100 kyr to few Myr old - including RXJ0720 at ${280_{-85}^{+210}}$ pc (Eisenbeiss 2011)
to $360 \pm 130$ pc (Kaplan et al. 2007); the other similar isolated neutron stars are probably more distant.

The Crab pulsar PSR~J0534+2200 has a characteristic age of 1240 yr as observed
from the pulse period (0.0331s) and its derivative ($4.2 \cdot 10^{-13}$ s/s, Lyne et al. 1993),
i.e. apparently close to the number of years since the AD 774/5 event.
If the true age of the Crab pulsar would be 1240 yr, one could conclude
that the breaking index was exactly $n = 3$ and that the initial spin period
was exactly 1 milli-second.
The Crab pulsar is of course associated with SN 1054, i.e. somewhat younger.
However, there are still several doubts as to whether the Crab SNR and/or pulsar
really formed in a core-collapse SN in AD 1054: 
(i) The expansion velocity of Crab SNR (e.g. van den Bergh 1973),
(ii) missing mass in the Crab SNR (Zimmerman 1998), 
(iii) the SN 1054 light curve possibly being not fully consistent
with a core-collapse SN (e.g. Collins et al. 1999), and 
(iv) possible early sightings of a very bright
source in April or May of AD 1054 (e.g. Collins et al. 1999), while the Chinese sightings
start on AD 1054 July 4. All those doubts would show that
the Crab SNR might have been from a SN Ia explosion and/or that the Crab SNR and/or pulsar
was not formed in AD 1054. However, we would like to note that all observables just mentioned
are still marginally consistent with a core-collapse SN and that it is highly dubious, whether the
sightings reported for April or May of AD 1054 are truely stellar events - apart from
the fact that their datings are highly uncertain (to up to a few decades), 
see e.g. Breen \& McCarthy (1995).
Given the characterisric age of the Crab pulsar of 1240 yr together with its
formation in AD 1054, it appears less problematic to conclude that either its breaking index
or its initial spin period were different from what is normally assumed.

\subsection{Strong flares on other stars}

The most energetic stellar flare was observed by ROSAT in X-rays 
on the young star YLW 15 in the $\rho$ Oph dark cloud ($\sim 119$ pc)
with energy output (over 5 hours) of $\sim 2 \times 10^{39}$ to $2 \times 10^{41}$ erg 
in X-rays (Grosso et al. 1997).
Even though the strong flare on YLW 15 may not have been a solar-type flare, but an
interaction with a circumstellar disk, let us estimate at which distance $d$ such a flare would have to happen
to be able to produce the AD 774/5 $\gamma$-ray event.
According to Equ. 1, a strong stellar flare with $E_{\rm event}=2 \times 10^{41}$ erg (Grosso et al. 1997)
of which a $\gamma$-ray flux of $E_{\rm obs} = 7 \times 10^{24}$ erg would be observed at Earth,
would have a distance $d \simeq 0.02$ pc or 3600 au;
this is independent of absorption, because $\gamma$-rays are effectivly unabsorbed.
If 4 to 6 times less energy would be needed (U13), then the
flare would need to happen at a $\sqrt 4$ to $\sqrt 6$ times larger distance,
i.e. at up to $d \simeq 0.05$ pc.
Of course, there is no star (except the Sun) at such small distances,
and in particular no young star with disk.
Hence, such a flare can be excluded as the cause for the AD 774/5 flare.
The Sun itself cannot produce a flare as on YLW 15, 
because there is no circumsolar gas disk left.
Even if we consider both the 4 to 6 times lower $^{14}$C and energy estimate (U13)
and strong beaming, i.e. another factor of 100 times lower total 
energy, then the distance of the young star flaring due to disk interaction would need to be $\simeq 0.5$ pc; 
there are no young stars with gas disks within such a small distance.

Strong flares were observed also on solar-type stars
with energies up to $2 \times 10^{38}$ erg in the optical as
for a flare on the G1V star S For (for 147 pc) on 1899 Mar 6 (Schaefer et al. 2000);
given that the new Hipparcos parallaxe of S For corresponds to 95 to 125 pc
within $1 \sigma$ (van Leeuwen 2007), the flare was slightly less energetic.
The {\em visually} observed flare of S For was suggested to be a mis-identification
(Payne-Gaposchkin 1952, Ashbrook 1959).
For a flare as large as supposedly observed on S For, it would have to happen at 114 au
to produce an energy input of $E_{\rm obs} = 7 \times 10^{24}$ erg at Earth,
if not beamed, but homogeneous.
It would have to happen at 255 au, if five times less energy would be needed (U13).
If another factor of 100 times lower total 
energy would be needed due to strong 
beaming, then the distance would need to be 2550 au.
Except for the Sun, there is no other star within this distance.
If S For is either younger than the
Sun or if it is a close binary or if is has an M-type companion, a large flare would
not be surprising, but also not be comparable to the Sun.
S For is known to be a binary star with $0.1$ to $0.3^{\prime \prime}$ separation between
1933 and 1991 (Mason et al. 2001 in the Washington Visual Double Star Catalog);
the secondary is 0.35 mag fainter in the optical (Mason et al. 2001) 
than the primary star (primary has spectral type G1V),
so that the secondary has probably a spectral type of mid-G;
the flare may have happened on the companion, which may not be a solar twin.
The next few strongest flares listed in Schaefer et al. (2000) did not happen
on solar-analog stars, but late F or mid G-type stars.

The rate of such large flares on other stars was very recently considered in Maehara et al. (2012)
based on Kepler observations of 365 superflares in 83000 stars:
According to their figure 2d, the largest flare recorded on a slowly rotating star with
5100 to 5600 K (i.e. cooler than the Sun) had an energy output of $\sim 6 \times 10^{35}$ erg and happens once every 
$\sim 5000$ yr
(no such strong flares were observed for slowly rotating stars with 5600 to 6000 K like the Sun).
From the same figure, we can also estimate the rate of flares (for slowly rotating stars with 5600 to 6000 K
like the Sun) with
$\sim 1.5 \times 10^{34}$ erg energy output 
(as needed for the AD 774/5 event due to $\gamma$-photons above 10 MeV from the Sun at 1 au distance)
to be once in about 1500 yr - or once every $\sim 750$ to 7500 yr as $1~\sigma$ error range (Maehara et al. 2012);
similar values are given for that type of stars in Shibayama et al. (2013), namely on flare in 800 to 5000 yr;
and the rate of flares with $\sim 2 \times 10^{35}$ erg energy output 
(as needed for the AD 774/5 event due to protons from in 1 au distance with the M12 energy estimate)
to be once in about 10,000 yr - or once every $\sim 4000$ to 400,000 yr (Maehara et al. 2012).
Hence, we will have one sufficiently large flare every $\sim 1500$ to 10,000 yr 
(or one event every 750 to 400,000 yr as full possible $1~\sigma$ error range).

If we consider the beaming of flares with $24^{\circ}$ opening angle, 
we would need 100 times less energetic flares.
Flares of that energy are not listed in Maehara et al. (2012).
According to Schrijver et al. (2012), such 100 times less energetic flares
are roughly 100 times more often than the larger ones quoted above,
i.e. once every few to $\sim 4000$ yr (by extrapolating beyond the observed range).
From detailed observations of recent decades and from the 11,000 yr old radiocarbon archive, 
an average frequency of one super-flare every few centuries can be excluded
(one every $\sim 4000$ yr would be possible).
On the other hand, if SPEs (on the Sun) are beamed with $\ge 24^{\circ}$ opening angle,
then flares on solar analog stars should also be beamed with similar angles. 
Hence, it might not be justified to apply a correction factor for beaming -
the Sun and solar-analog stars are assumed to be similar.

The above numbers from Maehara et al. (2012) would be consistent with roughly one event in 3000 yr, 
the age of the trees investigated by M12, or even one event in 11,000 yr (IntCal).

However, there are several sources of uncertainty in the Maehara et al. (2012) flare rates 
and energies: The flare rate and energies for solar-type slowly rotating stars is made up by
from only 14 stars, individual flare energies have uncertainties of $\pm 60~\%$,
and the number of such flares of solar-type slowly rotating stars with energies $\sim 10^{35}$ erg is only two (Maehara et al. 2012), 
namely KIC 10524994 and KIC 7133671 (Maehara, priv. comm.), hence very low-number statistics.
Those two stars were not investigated in more detail, yet, also not in Notsu et al. (2013),
where {\em sun-like} is defined as having an effective temperature of 5600 to 6000 K.\footnote{The three
stars investigated in detail in Notsu et al. (2013) are not {\em sun-like} according to the
definition in Notsu et al. (2013), but have lower temperatures.} Furthermore, the classification of the star
as either sun-like or solar-type or to belong to some temperature bin is also somewhat uncertain,
because these particular stars are faint and their temperatures were not determined by spectroscopy, but only
by multi-color photometry (G. Torres, priv. com.).

Furthermore, we have to consider that roughly half of the solar-type stars in the Galaxy are younger
than the Sun and that also half of them are binaries (Duquennoy \& Mayor 1991); both younger stars and
close binaries show more and stronger flares than the Sun. Then, the two stars with
the super-flares may have dMe-type companions, which would often show flares
(see e.g. Hambaryan et al. 2004).
There is a significant probability that one or both of those two stars with very large flares 
are younger than the Sun and/or have a close and/or M-type companion.

The 14 stars with strongest flares among stars with 5600 to 6000 K with slow
rotation periods (in Maehara et al. 2012) have rotation periods below 17 days, 
i.e. are all faster than the Sun. 
In figure 7 in Notsu et al. (2013) and figure 3 in Shibata et al. (2013), one can clearly
see that flares happen less frequently in stars with longer rotation periods,
in particular for rotation periods above some 10 days.
One of the two stars with the largest flares,
KIC 7133671, may show a periodicity of 3.25 days (Maehara et al. 2012, Maehara, priv. comm.),
which would indicate a much faster rotation (and younger age) than the Sun,
if this periodicity is interpreted as rotation period.\footnote{Since the amplitude of
its variation was only 18 ppm, which is smaller than the typical photometric precision, 
Maehara et al. (2012) have assumed that KIC 7133671 is a non-rapid rotator (Maehara, priv. comm.).}
Hence, the rates of large stellar flares given in Maehara et al. (2012) should be 
taken with care and should be regarded as upper limits when applied to the Sun.

Nogani et al. (2014) recently found for the two presumably solar-type Kepler super-flare
stars KIC 9766237 and KIC 944137 by comparing their spectroscopic rotational velocity and 
photometric rotational period that both are observed from nearly pole-on and concluded
that both would need to have polar spots with magnetic fields much higher than in the Sun.
Indeed, polar spots are otherwise not observed on solar-type stars and are considered to
exist only on stars which are either quite young (e.g. T Tauri stars, see e.g. Neuh\"auser et al. 1998)
or very magnetic (e.g. magnetic Ap stars, see e.g. Strassmeier 2009), or binary (e.g. RS CVn stars, see e.g. Vogt et al. 1999).

It was found recently that in the case of one of those two largest flares among 
presumable solar twins (as studied by Maehara et al. 2012) the extra photons during
the flare were emitted from a region outside the PSF of the main (presumable solar twin) star,
at a separation of 19 mas.
This means that the flare originated from another nearby (much fainter) star, either 
a faint low-mass companion or a faint background star (Kitze et al. 2014).
It follows that the rate of super-flares in sun-like stars is $\sim 1.6$ times smaller than claimed
in Maehara et al. (2012).

\section{Solar flares}

Let us now reconsider solar flares including SPEs (see Benz 2008 and Schrijver et al. 2012 
for reviews about solar flare observations).

According to Equ. 1, a flare with energy output $E_{\rm event}$ of $1.5 \times 10^{34}$ erg,
if happening on our Sun ($d=1$ au), would produce an energy input of $E_{\rm obs} = 7 \times 10^{24}$ erg at Earth,
the energy observed in AD 774/5 - in case that the original source for the $^{14}$C and $^{10}$Be 
production by thermal neutrons were $\gamma$-photons above 10 MeV with a SN-like spectrum with a power-law index
of $-2.5$ (M12).
In case that protons from a SPE would have been the original source for the $^{14}$C and $^{10}$Be
production, one would need a proton energy of $8 \times 10^{25}$ erg at Earth or
$2 \times 10^{35}$ erg at the Sun (M12).

Such strong super-flares or SPEs were never observed on the Sun (Jackman et al. 1995). 
The largest flare observed on the Sun (on 1859 Sept 1) had an energy output of $\sim 10^{32}$ erg
(Carrington 1859, Hodgson 1859, Tsurutani et al. 2003, Townsend et al. 2006),
i.e. by a factor of $\sim 2000$ too weak to explain the 
AD 774/5 event as SPE. A SPE with $10^{29}$ to $10^{32}$ erg (Baker 2004, Baker et al. 2004) would have been too weak, too.
Even if 4 to 6 times less $^{14}$C was produced and, hence, 4 to 6 times less energy was needed
(U13), then the Carrington flare was still $\sim 400$ times too weak.

If the flare would be beamed (e.g. to an angle of only $\sim 24^{\circ}$), 
100 times lower 
energy would be needed (Melott \& Thomas 2012). 
An event as strong as the Carrington event would still be $\sim 20$ times too weak (for the M12 energy estimate),
or only $\sim 4$ times too weak (for the U13 energy estimate).
Now, in the AD 774/5 event, the $^{14}$C 
(to $^{12}$C ratio) increased (on several data points) by $1.2~\%$ (with a mean error bar of $\pm 0.12~\%$), 
in total a $7~\sigma$ signal (M12, U13).
If the AD 774/5 event was only 4 times more energetic than the AD 1859 Carrington event,
then a 4 times lower signal in $^{14}$C (and $^{10}$Be) should be detectable in the AD 1859 data
over several years.
However, neither $^{14}$C nor $^{10}$Be peaks were found for AD 1859, recently confirmed by M13 for $^{14}$C,
see also our Fig. 2, where the $^{14}$C even decreases after AD 1859 (due to the Schwabe cycle phase). 
Hence, the AD 774/5 event (as solar flare) either was not beamed that strongly,
and/or it would have been much more than 4-6 times stronger than Carrington,
and/or the lower energy estimate (Usoskin et al. 2013) is not correct,
and/or such solar flares cannot form (enough) $^{14}$C and $^{10}$Be,
and/or the Carrington event itself was much softer than the AD 774/5 event.
A similar conclusion was previously drawn by Kocharov et al. (1995), Stuiver et al. (1998b), 
and Usoskin \& Kovaltsov (2010).

However, even if the AD 774/5 event was as hard as the hardest solar flare observed in 
the last century (SPE 1956 Feb 23), it would not have been hard enough for the $^{14}$C and $^{10}$Be
production observed in AD 774/5 (Cliver et al. 2014).
For SPE 1956, neither $^{10}$Be nor cosmic ray peaks 
were observed, see Fig. 3: Cosmic ray data from Huancayo and Climax have vertical cutoff rigidities of $\sim 13$
and $\sim 3$ GV, respectively, so that they are relevant for our study, namely including those
particles which could possibly form $^{14}$C and $^{10}$Be.
We do not see any increase in cosmic rays - neither in Huancayo nor Climax - 
immeadiately after SPE 1956, i.e. in Feb 1956, when the SPE has hit Earth,
neither in the yearly nor monthly data sets.
However, we do see possible small {\em drops} in May 1956 in cosmic rays 
(and a broader depression in cosmic rays from about April to July 1956) 
in both Huancayo and Climax, i.e. a few month after the SPE,
which might due to stronger modulation of cosmic rays by a more active Sun.
The Forbush delay between solar activity change and cosmic ray response on Earth is 6 to 12 month (Forbush 1954),
it is shorter (only a few month) during A$+$ cycles as in 1956 (until the magnetic reversal during the sunspot maximum in 1957/58).
We should restrict the analysis to data with 1-yr time resolution,
because $^{10}$Be ice core (and $^{14}$C tree ring) data also have time resolution of 1 yr (and not better).
We also consider the two $^{10}$Be records available for 1956, namely both NGRIP and Dye-3,
both with 1 yr time resolution, none of them shows any signal after SPE 1956.
The $^{14}$C record is not useful anymore for the time around 1956 because of the Suess and atomic bomb effects.

Hence, we can add to the arguments given in Cliver et al. (2014) that the very hard SPE 1956 was also not
detected in cosmic rays nor in $^{10}$Be, which makes it even more dubious that the AD 774/5 event would 
have been detectable in such data. If there was any effect after SPE 1956, then a drop in cosmic rays,
possibly due to a more active sun, i.e. stronger modulation, but no increase in cosmic rays.

The rate of such large flares on the Sun (including SPEs) was considered in Schrijver et al. (2012):
The rate of solar flares with $\sim 1.5-20 \times 10^{34}$ erg energy output 
(as needed for the AD 774/5 event in 1 au distance) was found to be once every 
$\sim 1000$ to 10,000 yr (their figure 3). 
Flares with 100 times lower total energy (including those beamed with $\sim 24^{\circ}$) happen 100 times more frequent,
i.e. once in $\sim 10$ to 100 yr. All these numbers depend strongly on how to extrapolate
from smaller flares to larger flares, because such large flares were never observed on the Sun,
so that they are highly uncertain. Both rates are roughly consistent with Maehara et al. (2012).
A rate of once in $\sim 10$ to 100 yr can be excluded from solar observations of the last decades.

Schrijver et al. (2012) also argue that spots connected to flares with energy of $\sim 10^{34}$ to $10^{35}$ erg
would cover $\sim 12$ to $40~\%$ of the surface of the Sun, so that they are pratically impossible:
The energy of a flare and the sizes of sunspots are limited due to the solar magnetic field, 
which itself is limited by photospheric pressure equilibrium to a few kG - hence, the limits given in Schrijver et al. (2012).
Neither such large flares (aurorae) nor sunspots were ever recorded in the last few millennia, even though many
naked eye sunspot and aurorae observations were done, e.g. Fritz (1873) and Clark \& Stephenson (1978), 
so that such very large sunspots or very strong aurorae would have been detectable.
Very young stars can produce such large flares, because they rotate faster and, hence,
have larger magnetic fields (e.g. Preibisch et al. 1995, 1998; Neuh\"auser et al. 1998, 2009).
Furthermore, as argued by Shibata et al. (2013), such spots which might be able to power such large
super-flares (up to $10^{35}$ erg) would last for $\sim 10$ years, which was clearly never observed
for the last two millennia, but such large and long-lasting spots would have been observable 
(even though only through clouds and/or near the horizon and/or shortly after a volcano eruption,
when sufficient dust blocks some sunlight). Also, during the Carrington event, the spot was not seen
for a unusual long period of time (Carrington 1859, Hodgson 1859, Tsurutani et al. 2003).

Shibata et al. (2013) argue that such super-flares may happen after a long grand minimum, i.e. after
the Sun has strored enough magnetic energy for a super-flare; however, they also mention some further 
theoretical problems related with super-flares, e.g. an energy budget problem and the problem of
energy diffusion. We also note that no particular large solar flares were observed after the
grand minima in the last millennium.
According to figure 1 in Shibata et al. (2013), flares as large as the Carrington event with $10^{32}$ erg
would happen almost every year, which is clearly rejected by observations of the last centuries.

The fact that such large flares were observed on some apparently solar-type stars with similar
temperature (Maehara et al. 2012) therefore seems contradictory, since they should have similar radii 
and magnetic fields as the Sun.
However, it is possible that the two largest flares observed by Maehara et al. (2012) were observed
on faster rotating (i.e. younger) solar-type stars and/or on (as yet unknown) binaries
(e.g. close binaries or with M-type companions), where flares are much more violent and frequent (e.g. Hambaryan et al. 2004).
Indeed, Kitze et al. (2014) found that another star near one of the two presumable solar twin stars
with the largest flares in Maehara et al. (2012) was responsible for the flare (and not the
presumable solar twin star), either a companion or a background star.

We should then consider that the stellar flares observed by Maehara et al. (2012) with the Kepler satellite
were observed in the optical band, while the AD 774/5 event was due to $\gamma$-rays or cosmic rays:
E.g., a flaring plasma confined in a loop with a very hot plasma temperature T=$10^{9}$ K 
(the observed range is 1 to $20 \times 10^{7}$ K),
a plasma density of $n_{e}=10^{12}$ cm$^{-3}$ 
(the observed range is $10^{10}$ to $5 \times 10^{11}$ cm$^{-3}$) 
and a plasma loop volume V = $10^{34}$ cm$^{3}$ 
(the estimated range is $2 \times 10^{29}$ to $7 \times 10^{32}$ cm$^{3}$),
corresponds to the enormous emission measure of $10^{58}$ cm$^{-3}$
(observed typical values are in the range of $10^{51}$ to $10^{55}$ cm$^{-3}$,
see e.g. Reale 2007),
at one light year distance from the Earth, the incident
$\gamma$-ray energy above 1~MeV is $\sim 7 \times 10^{19}$ erg of persistent
emission during one year, several orders of magnitude less than
needed for the production of radiocarbon observed in tree rings at AD 774/5.

In summary, the rate of both stellar and solar flares with the neccessary energetics may roughly 
be consistent with zero or one event in 3000 yr or 11,000 yr, 
as observed for the AD 774/5 $^{14}$C variation (M12),
so that we will continue to consider them.

\section{Discussion}

Given the (still highly uncertain) rate of large flares on solar analog stars, we estimated whether a solar super-flare
as large as needed for the AD 774/5 event may be possible within some 3000 yr.
Such a rate of large flares on solar analog stars is an average of all
stellar activity phases (or even valid only for maximum activity in grand maxima).

Above (Sect. 2), we concluded to expect at most one sufficiently (for the AD 774/5 energetics) large
solar flare every $\ge 1500$ to 10,000 years (or one every $\ge 750$ to 400,000 years as full
$1~\sigma$ error range). This translates to a probablity for one flare within 3000 years
of as low as $\sim 0.3$ (or $0.008$ considering the full range).

Also, according to Kovaltsov \& Usoskin (2013), considering data in lunar rocks,
the occurence rate of such large solar super-flares is smaller than $10^{-4}$ per yr.

This probability for a large solar flare can be compared to the probability
for, e.g., a short GRB beamed towards Earth: 
$0.0013$ to $0.0005$ for short GRBs (within 4 kpc within 3000 yr), 
$\le 0.075$ for mergers of two neutron stars as short GRB beamed towards Earth (within 4 kpc within 3000 yr),
or $0.04$ to $0.20$ for mergers of two White Dwarfs as short GRB beamed towards Earth (within 4 kpc within 3000 yr),
also $1~\sigma$ error ranges, see HN13 for details and references.

If we consider only one such event within 11,000 yr, as shown by Usoskin \& Kovaltsov (2012) 
based on $^{14}$C data, we would expect a probability of $0.15$ to $0.73$ for a merger of two 
White Dwarfs as short GRB beamed towards Earth within 4 kpc.

If we consider the revised $^{14}$C production ratio and energy input to Earth for AD 774/5 (U13),
i.e. 4 to 6 times less energy, then the event, as e.g. short GRB, would be a factor of $\sqrt 4$ to $\sqrt 6$
more distant, i.e. at up to $\sim 10$ kpc. Then, the probability for such an event, e.g. a short GRB
(or a merger of two compact objects), is also larger by a factor of $\sqrt 4$ to $\sqrt 6$.

Hence, the probability ranges of very large solar flares and short GRBs overlap.
Short GRBs can happen due to the merger of compact objects like neutron stars and Black Holes (BH),
maybe also White Dwarfs. The merger rate of two Black Holes is up 1 to 1000 mergers per Myr
per Milky Way Equivalent Galaxy (Kalogera et al. 2004). 
The latter upper limit rate corresponds to one BH-BH merger per kyr. 
If such a merger would be observable as short GRB, one would have to correct the rate for the
beaming fraction $f=0.01$ to 0.13 for short GRBs (Rezzolla et al. 2011).
For $f=0.13$, one would then expect up to one merger in $\sim 7.7$ kyr
(within $1~\sigma$ error bars), pointed towards Earth as short GRB,
from anywhere in the Galaxy. With the recently revised estimate of the $^{14}$C and energy content
of the AD 774/5 event (U13), a short GRB would need to take place within some $\sim 10$ kpc,
i.e. in a volume including the Galactic Center. The expected rate of up to
one merger in $\sim 7.7$ kyr in the whole Galaxy would correspond to a few 
up to one merger in $\sim 10$ kyr for a volume within $\sim 10$ kpc.
This rate is only a factor of a few deviant from the 3 kyr age of the Japanese trees.
Given that no such other strong or stronger event was found in 11 kyr $^{14}$C IntCal
data (the AD 993/4 event was weaker), the rates of BH-BH mergers and strong $^{14}$C
events may even be consistent with eachother.

The rate of $^{14}$C increases as the one in AD 774/5 (M12) is of course also highly uncertain,
as only one is observed in 3000 yr (or 11,000 yr), i.e. very small-number statistics
(2 flares, i.e. $2 \pm \sqrt 2$ = $2 \pm 1.4$ flares):
With the observed single large event within 3000 yr (or within 11,000 yr, respectively), 
the $68.3~\%$ confidence interval or credibility range from the Bayesian perspective 
(Love 2012) for the rate is $0.000083$ to $0.00075$ large events per yr 
(i.e. $0.25$ to $2.25$ large events within 3000 yr),
or $0.000023$ to $0.000205$ large event per yr, obtained from one large event within 11,000 yr
(i.e. again $0.25$ to $2.25$ large events within 11,000 yr) - assuming that solar flaring rate can be described with
stationary Poissonian process independent from the activity phase of the Sun.
However, it is clear that solar flare rate is changing during and with the Schwabe cycle 
(and also with longer cycles), i.e. it must be described by a non-stationary Poissonian process
(see, e.g. Wheatland et al. 1998, Gorobets \& Messerotti 2012). 
Indeed, by analysing the flare waiting time distribution, based on a GEOS X-ray/H$\alpha$
flares of the Sun, it turns out that average waiting time during
solar minimum is more than one order of magnitude longer in comparison
to the maximum, which is also true for sunspot numbers (Gorobets \& Messerotti 2012). 
Thus, given the tight correlation between the average sunspot numbers 
and flare activity, our linear scaling is justified for the probability estimate.

If we consider that the observed rate of events at least as large as the AD 993/4 event is
one in 1130 yr, 
than the numbers are as follows: For one event in 1130 yr, the event rate
is 0.00044 to 0.00040 per year (within $68~\%$ confidence); for two events in 1130 yr,
the event rate is 0.00061 to 0.00026 per year (again within $68~\%$ confidence).

\begin{figure*}
\begin{center}
{\includegraphics[angle=270,width=17cm]{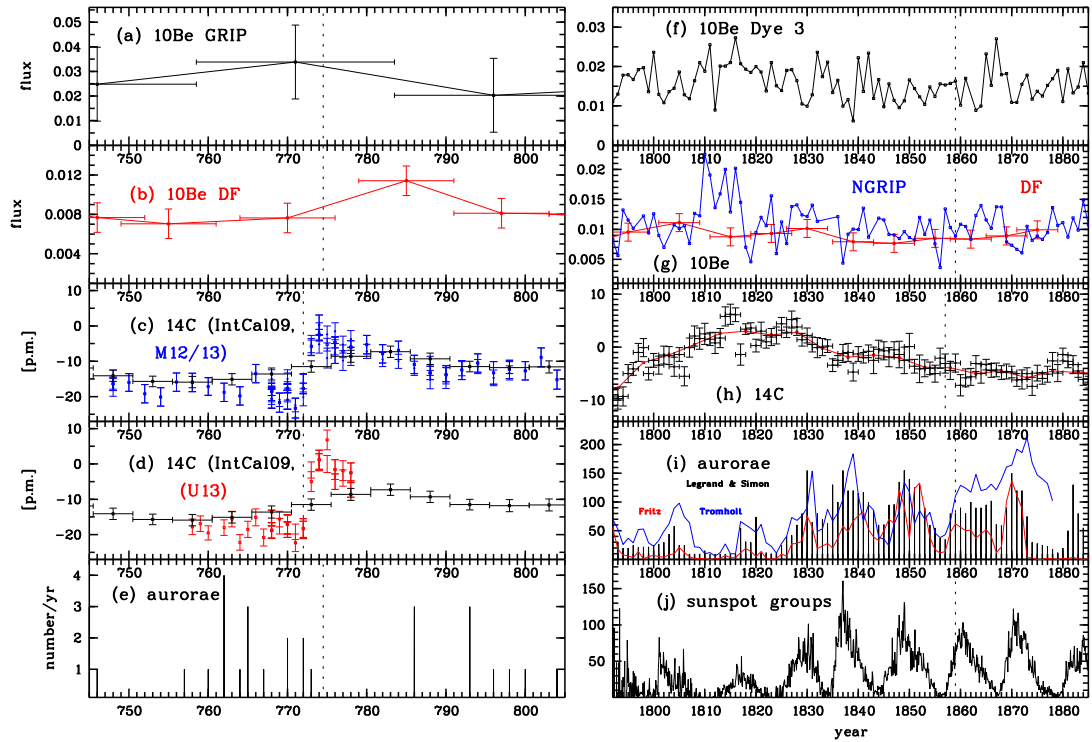}}
\caption{Historic records around the $^{14}$C variation in AD 774/5 from AD 745 to 805
(left) and around the AD 1859 Carrington event from 
AD 1792 to 1885 (right):
(a) $^{10}$Be flux in atoms cm$^{-2}$ s$^{-1}$ for Greenland GRIP data (Vonmoos et al. 2006, recieved from J. Beer, priv. comm.);
(f) $^{10}$Be flux in atmos cm$^{-2}$ s$^{-1}$ for north pole Dye-3 Swiss Ice core data (Beer et al. 1990,
recieved from J. Beer, priv. comm.);
(b) and (g) $^{10}$Be flux in atoms cm$^{-2}$ s$^{-1}$ for south pole Dome Fuji data in red (Horiuchi et al. 2008);
in panel (g) also NGRIP $^{10}$Be flux in blue (Berggren et al. 2009a,b);
(c) $^{14}$C with 1- to 2-yr time resolution (M12 and M13 data in blue,
while the full black line with data points with error bars show the IntCal09 data with lower (5 yr) time resolution;
(h) $^{14}$C with 1-yr time resolution (Stuiver et al. 1998b) in black and from IntCal09 in red;
(d) $^{14}$C with 1- to 2-yr time resolution (U13, blue) and IntCal09 in black;
(i) number of aurora sightings per year from Fritz (1873) as black lines,
Legrand \& Simon (1987) as red line, and from
Tromholt (1902) for Skandinavian aurorae (in blue), 
plotted are number of days per year with sightings, so that the Carrington event itself
does not show up significantly, where there were tens of sightings within a few days
(the blue line shows the extra aurora peak at around 1795 indicating
the extra Schwabe cycle, i.e. two Schwabe cycles inside the old cycle 4);
(e) number of aurora sightings per year (from Neuh\"auser \& Neuh\"auser 2014b);
(j) sunspot group numbers from Hoyt \& Schatten (1998).
In panels (c), (d), and (h), we plotted the $^{14}$C data $\sim 2.5$ yr ahead of their measurement
time, because they were produced at about that time due to the carbon cycle.
The vertical lines indicate the AD 774/5 years (left), and the AD 1859 year of the Carrington event (right);
also 2 yr earlier for $^{14}$C.
The $^{14}$C data (parts c and d, M12, M13 and U13) show the strong fast rise from AD 774 to 775. 
The Carrington event was not detected in $^{14}$C nor in $^{10}$Be,
not even if the $^{10}$Be would need to be shifted by a few years due to timing precision.
The right panels also shows the so-called Dalton minimum from about AD 1793 to 1825.}
\end{center}
\end{figure*}

If the M12 energy estimate for the AD 774/5 event is scaled down by both a factor of 100
due to beaming (Melott \& Thomas 2012) and another factor of 4 to 6 (U13),
then a solar flare as cause for the AD 774/5 $^{14}$C increase would need to be only $\sim 4$
times larger than the AD 1859 Carrington event (for similar spectra). 
As we can see in our Fig. 2, the Carrington event\footnote{ 
While the energy estimate for the Carrington event is only very rough and while Carrington himself noted 
that {\em one fly does not make a summer} regarding this event, we would like to remark that, even if there 
would have been larger flares in the last $\sim 150$ yr, they are not detected in $^{14}$C nor $^{10}$Be data with 1-yr time-resolution.}
was not detected in $^{14}$C nor in $^{10}$Be,
even though much more precise data with one-yr time resolution are available. There is not even only
a small increase. It was noticed before by Usoskin \& Kovaltsov (2012) that their model
is not consistent with the AD 1859 Carrington event data (their figure 4).

If a solar flare or SPE would hit the northern and southern polar regions on Earth with the same energy, 
then the same amount of $^{10}$Be production would be expected.
According to Usoskin \& Kovaltsov (2012), the AD 774/5 event is not detected
in the GRIP data yielding an upper limit in the fluence above 30 MeV
of $F_{30} \le 3 \times 10^{10}$ photons/cm$^{2}$ at the 0.03 significance level for Greenland.
Even though the $^{10}$Be at the northern and southern poles are not always identical,
non-detection of $^{10}$Be at the northern pole may be seen as evidence against a solar flare (or SPE):
U13 state that their scenario (of a presumable large solar flare) 
{\em yields an expected peak that is higher (by about $2~\sigma$) than the observed peak.
Thus, the GRIP series is not fully consistent with our [U13] scenario and the other
data series, but the existence of the peak cannot be excluded at the $5~\%$ level} (U13).
Regarding the southern Dome Fuji data, U13 write that the AD 785 {\em peak is
delayed by several years} compared to their scenario and that their data have to be {\em shifted by 5 years
to match the observed data} (U13, all in their section 4 and figure 3).
We would like to note, as seen in Fig. 2, that there is an increase in $^{10}$Be GRIP data for
the time interval AD 760-782 (compared to the time before and after that interval),
while the $^{10}$Be increase in Dome Fuji data is seen in the time interval AD 780-790;
none of them is highly significant. The difference in timing might be attibuted to the
low precision in the absolute timing of $^{10}$Be data.

\begin{figure*}
\begin{center}
{\includegraphics[angle=270,width=17cm]{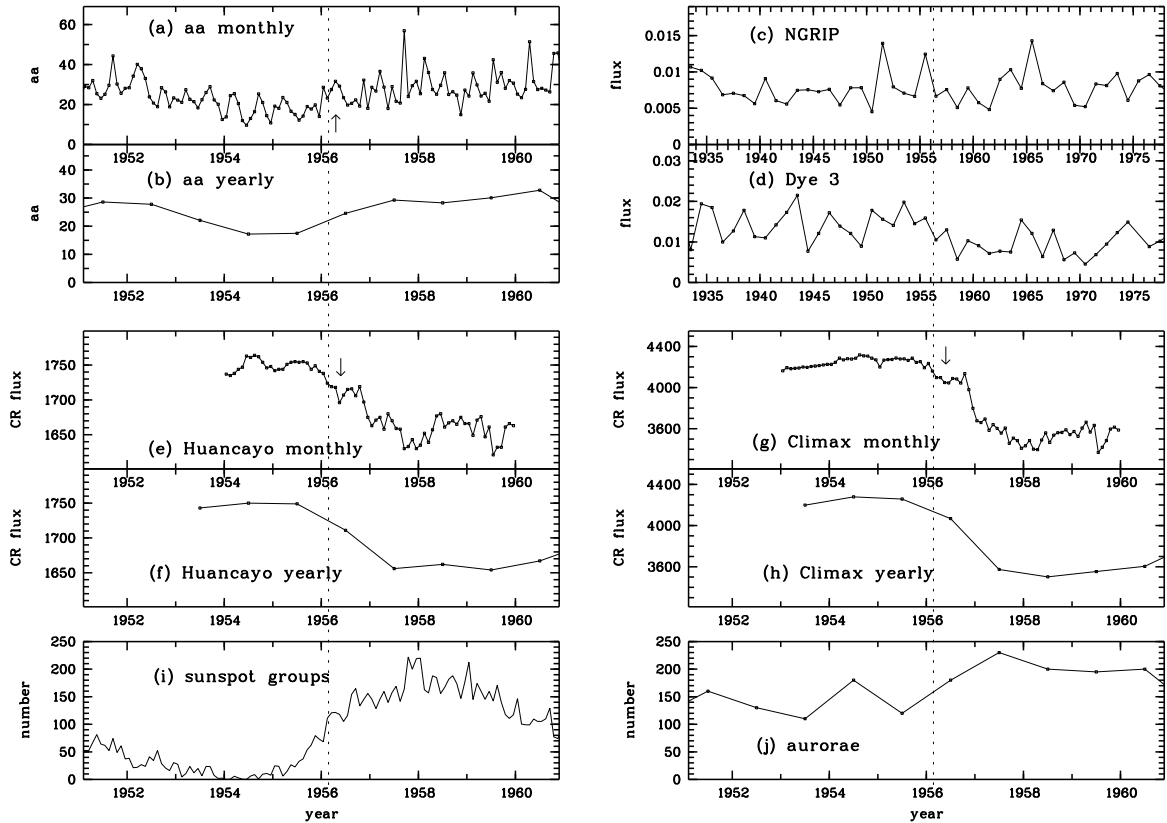}}
\caption{Solar activity proxies around the Solar Proton Event SPE 1956 on AD 1956 Feb 23 (from AD 1951 to 1960):
(a) Monthly geo-magnetic aa-index (in nT from ftp.ngdc.noaa.gov/STP/SOLAR$\_$DATA/).
(b) Yearly geo-magnetic aa-index.
(c) NGRIP $^{10}$Be flux in atoms cm$^{-2}$ s$^{-1}$ from Berggren et al. (2009ab).
(d) Dye-3 $^{10}$Be flux in atoms cm$^{-2}$ s$^{-1}$ from Beer et al. (1990), 
received from J. Beer in electronic form.
(c \& d) Note that we plot a different y-axis, 
namely from AD 1933 to 1978, i.e.
a larger range, because $^{10}$Be timing is much more difficult; however, there is no strong
peak, not even within $\pm 20$ years around SPE 1956.
(e) Monthly cosmic ray flux from Huancayo (ftp.ngdc.noaa.gov/STP/SOLAR$\_$DATA/).
(f) Yearly cosmic ray flux from Huancayo.
(g) Monthly cosmic ray flux from Colorado/Climax (ftp.ngdc.noaa.gov/STP/SOLAR$\_$DATA/).
(h) Yearly cosmic ray flux from Colorado/Climax.
(i) Monthly sunspot group numbers from Hoyt \& Schatten (1998).
(j) Yearly aurorae from Legrand \& Simon (1987).
In panels a, e and g, small arrows point to an increase in aa (panel a) and
a small drop in cosmic rays at 1956.4 (1956 May) and/or a broader depression in cosmic rays 
from about April to July 1956 (panels e and g) - maybe due to stronger modulation by a more active sun
(the time delay after the SPE on 1956 Feb 23 is due to the Forbush effect (Forbush 1954)
being 
a few month in the A$+$ cycle in 1956).
A short-term strong rise is seen in aa already from Dec 1955 to Jan 1956 (too early for the SPE 1956), 
in Feb 1956 there is no strong aa signal. The rise in aa from Dec 1955 to about March 1955 shows
that solar activity and wind were larger for that time, so that less cosmic rays came in (hence, the depression).
Given that both $^{10}$Be in ice and $^{14}$C in trees have only 1-yr time resolution,
we should consider the {\em yearly} cosmic ray and aa data, then we see no signals at all.
In panels c and d, we see that there is also no signal from SPE 1956 in $^{10}$Be data.
The downward trend of the cosmic rays and of $^{10}$Be from AD 1955-1959
is explained by the rise in solar activity from AD 1955-1959, i.e. solar modulation.
$^{14}$C is not available anymore 
for the time around 1956 (since about AD 1900) 
due to the Suess and bomb effects.}
\end{center}
\end{figure*}

Even if such a solar flare or SPE several times larger than the Carrington event in AD 1859
would have happened in AD 774/5, it would have resulted in strong auroral sightings down to the tropics. 
For the smaller Carrington event in AD 1859, Fritz (1873) lists more than 50 aurora sightings 
for the northern hemisphere (even on Hawaii at $b=+20^{\circ}$ and 
on the {\em Barke Baltimore} at $b=+14^{\circ}$ northern latitude)
and also several dozens for the southern hemisphere.
However, there are no such records for AD 774/5.
There are, however, a few records on aurorae within $\pm 10$ yr around AD 774/5 (but not in AD 774/5),
showing that aurorae in those decades were noticed and that records do exist (but no excess).
As shown in Neuh\"auser \& Neuh\"auser (2014b), there were aurora observations
in Arabia in AD 793 and 817, i.e. a quite southern location, hence strong aurorae -
but without another $^{14}$C increase at those times (Neuh\"auser \& Neuh\"auser 2014b).

While M12 and Schaefer et al. (2000) argued that such a strong event would have resulted
in ozone layer destruction, and, hence, an extinction level event on Earth, this was recently
challenged by Thomas et al. (2013): In their more detailed calculations, they show that
the UVB radiation would have increased, but reducing the ozone layer by only some $10~\%$,
so that there would have been no extinction level event; however, erythema and possibly skin cancer
would have increased by 14 to $160\%$ depending on the strength and hardness of the flare;
the largest effect would have been expected at $b=55^{\circ}$ latitude (Thomas et al. 2013).
There are no indications in medieval reports found so far about such effects beginning just after 
the event, i.e. in AD 774/5 - even though such reports are in many cases very detailed. 
Thomas et al. (2013), however, may have over-estimated the effect: They used a fluence of
$3 \times 10^{10}$ protons cm$^{-2}$ {\em as a lower bound} for all three cases they calculate,
while Usoskin \& Kovaltsov (2012) already noticed that {\em a fluence greater than
$3 \times 10^{10}$ protons cm$^{-2}$ is inconsistent at the 0.03 significance level},
namely inconsistent with non-detec- tion of $^{10}$Be on Greenland (Usoskin \& Kovaltsov 2012).

Even if the aurorae were not noticed or not reported or if the reports were not yet found,
we can consider whether the $^{14}$C than $^{10}$Be production is consistent with a solar event.
According to M12, a solar flare or SPE can be excluded for the AD 774/5 $^{14}$C event,         
because solar flare are not hard enough. In HN13, we have estimated that in the AD 774/5 event,
at least $270 \pm 140$ times more $^{14}$C than $^{10}$Be was produced; here we assumed that all
the $^{10}$Be detected for the decade around AD 774/5 was produced during one year, namely the
same year as the $^{14}$C in AD 774/5; the ratio is an upper limit, because some of the 
$^{10}$Be produced in that decade may have been produced by other events.
According to detailed calculations by Usoskin et al. (2006), SPEs can produce only 25 to 27
times more $^{14}$C than $^{10}$Be (their table 1), while Usoskin \& Kovaltsov (2012) have shown 
that SPEs can produce 38 times more $^{14}$C than $^{10}$Be (their section 2.2).
Both values would be too low for the AD 774/5 event.

U13 suggest that 4 to 6 times less $^{14}$C was produced in AD 774/5.
Given that the amount of $^{10}$Be production should remain unchanged, the differential
production ratio of $^{14}$C to $^{10}$Be would then be reduced to $\ge 54 \pm 30$
(scaling from HN13), which is then marginally consistent with the expectation from
Usoskin et al. (2006) and Usoskin \& Kovaltsov (2012), namely 25 to 38 times more $^{14}$C than $^{10}$Be.

Summary: There are several problems with the interpretation of the AD 774/5 $^{14}$C variation as solar super-flare: 
\begin{enumerate}
\item Neither exceptionaly strong aurorae nor any sunspots were observed around AD 774.
\item The differential production ratio of $^{14}$C to $^{10}$Be may not be consistent with a solar flare.
\item If the AD 774/5 $^{14}$C event would have been a solar super-flare only a few times larger than the
Carrington event (U13), and having similar spectra, then the Carrington event should have been 
detected in $^{14}$C and/or $^{10}$Be variations, which is not the case (Fig. 2).
Hence, the AD 774/5 event either was not a solar super-flare or it was much more than 4 times stronger than Carrington
(in the latter case, aurorae should have been observed, which is not the case).
\item Strong solar flares were never detected before as strong $^{14}$C or $^{10}$Be incresases in data with one year time resolution.
\item The rate of strong super-flares on solar analog stars (averaged over all activity
phases or obtained only in strong activity phases) is highly uncertain and very low, possibly zero;
the rate of very hard flares (needed for the production ratio of $^{14}$C to $^{10}$Be) is even more
uncertain and smaller.
\item The only two early-G type Kepler stars observed to have shown a large super-flare (Maehara et al. 2012)
may be either fast rotators or binaries, i.e. not really solar twins; 
for one of them, it was recently found
that another star (companion or background) next to the presumable solar twin produced the flare (Kitze et al. 2014).
\item If the AD 993/4 event would also need to be a solar super-flare, too, the problem of the flare rate
is even more severe: One would need an even larger flare rate.
\item By comparing energetics and spectrum of the very hard 1956 SPE with AD 774/5, 
Cliver et al. (2014) found strong doubts on the solar flare interpretation for AD 774/5;
SPE 1956 also was not detected as spike in cosmic rays; there are also no
variations due to the flare in cosmic ray nor $^{10}$Be data with 1-yr time resolution (Fig. 3).
\item It is dubious, whether the Sun itself can produce such a large super-flare, given its magnetic field (Schrijver et al. 2012).
\end{enumerate}
Thus, both the proposed causes for the observed short-term increases of $^{14}$C or $^{10}$Be in tree rings or ice cores, 
a large SPE and a Galactic short GRB, face certain difficulties like low event rates. \\
A solar super-flare is also questionable unless some new physics is developed that could explain how solar energetic particle
irradiation can be better focussed by a factor of 1000 and/or local accelaration near Earth. This implies that the
solar super-flare interpretation does not fit within the current knowledge of solar and stellar flares.
Also, if Usosk- in et al. (2014) is correct in concluding that, in the last few decades of the last century,
the Sun were in its highest activity mode for the last few millennia, would it not be problematic that the strongest solar 
flares did not occur in the most recent time, i.e. during the presumable recent Grand Maximum~?

\acknowledgements
We used the ATNF online catalog of pulsars maintained by G.B. Hobbs and R.N. Manchester,
and the McGill online catalog of SGRs and AXPs maintained by the McGill Pulsar Group.
We are grateful to H. Maehara for further information about the Kepler flare stars in his 2012 Nature paper.
We retrieved the $^{14}$C IntCal09 and $^{10}$Be Dome Fuji records from www.radiocar- bon.org and ftp.ncdc.noaa.gov, respectively.
We got the $^{10}$Be GRIP and Dye 3 data in electronic form from J. Beer.
We retreived further $^{10}$Be from the IGBP PAGES/World Data Center for Paleoclimatology.
The $^{14}$C data plotted in Usoskin et al. (2013) for the German Oak tree were sent to us in electronic form by L. Wacker.
We thank all those for their electronic material.
We also acknowledge good comments from an anonymous referee.
RN would like to thank K. Kokkotas, Y. Eksi, A. Benz, G. Torres, and D.L. Neu- h\"auser for discussion.
We would like to thank the German national science foundation (Deutsche Forschungsgemeinschaft, DFG) 
for financial support in the collaborative research center Sonderfor- schungsbereich SFB-TR 7 
Gravitational Wave Astronomy project C7.

\end{document}